\documentclass[column,showpacs,preprintnumbers,amsmath,amssymb]{revtex4}
\preprint{}
\usepackage{graphicx}
\usepackage{dcolumn}
\usepackage{bm}
\usepackage{mathrsfs}
\begin{document}
\title{Michelson interferometer for measuring temperature}
\author{Dong  Xie}
\email{xiedong@mail.ustc.edu.cn}
\affiliation{Faculty of Science, Guilin University of Aerospace Technology, Guilin, Guangxi, P.R. China.}

\author{Chunling Xu}
\affiliation{Faculty of Science, Guilin University of Aerospace Technology, Guilin, Guangxi, P.R. China.}

\author{An Min Wang}
\affiliation{Department of Modern Physics , University of Science and Technology of China, Hefei, Anhui, China.}

\begin{abstract}
We investigate that temperature can be measured by a modified Michelson interferometer, where at least one reflected mirror is replaced by a thermalized sample. Both of two mirrors replaced by the corresponding two thermalized samples can help to approximatively improve the resolution of temperature up to twice than only one mirror replaced by a thermalized sample.  For further improving the precision, a nonlinear medium can be employed. The Michelson interferometer is embedded in a gas displaying Kerr nonlinearity. We obtain the analytical equations and numerically calculate the precision with parameters within the reach of current technology, proving that the precision of temperature can be greatly enhanced by using a nonlinear medium. Our results show that one can create an accurate thermometer by measuring the photons in the Michelson interferometer, with no need for directly measuring the population of thermalized sample.
\end{abstract}

\maketitle

\section{Introduction}
 Creating accurate thermometers is very significant in understanding thermodynamics and developing new quantum technology.
 As we all know, the Boltzmann constant $K$ is less accurate than the other basic constants except the gravitational constant.
 Improving the detection precision is necessary and important.  And it will bring breakthroughs in other fields, such as medicine, biology, and material science\cite{lab1}.

Most of physicists consider to directly measure the energy level population of sample to be tested\cite{lab2,lab3,lab4}.
 In the reference\cite{lab5}, the authors consider that thermometry may be mapped to the problem of phase estimation. At last, the temperature is still measured by the population. In one word, it is necessary to touch the sample directly for higher resolution of temperature.

In this article, we consider that an optical method is used to measure the temperature with no need for direct contact with the sample. A modified Michelson interferometer can be an accurate thermometer.  Namely, the mirrors in the Michelson interferometer are replaced by the thermalized samples. The information of temperature is encoded in the photons reflected by the sample. The temperature of sample can be obtained by measuring the reflected photons.

We compare the difference between one and two thermalized samples. In general, replacing both of two mirrors by thermalized samples performs better than only replacing one mirror with a thermalized sample.
For further enhancing the precision, nonlinear medium becomes an optimal choice\cite{lab6}.  The Michelson interferometer is embedded in a gas with Kerr nonlinearity. Nonlinearity is an useful resource, which allows us to get better resolution than linear ones. A lot of works have proved it \cite{lab7,lab77,lab777,lab7777,lab78,lab79,lab70} in different physical conditions.

The rest of this article is arranged as follows. In Section II, we briefly introduce the modified  Michelson interferometer and the measurement method. In section III, both of two mirrors are replaced by the themalized samples.
In Section IV, nonlinear medium is employed to enhance the precision beyond Hensiberg limit. We numerically calculate the uncertainty of temperature with parameters within the reach of current technology in Section V. In Section VI, we discuss about the efficiency of detection and reflection, a numerical calculation  of the resolution, as well as a simple analysis of more complex situation.   A concise conclusion is presented in Section VII.
\section{Modified Michelson interferometer}
A modified Michelson interferometer is used to measure the temperature of sample as shown in Fig.1. Here, one mirror is replaced by the sample. In the absence of the fluctuation of sample, the two arms are equal, $L_1=L_2=L$. Due to the thermal fluctuation of sample, the length of arm $L_2$ is described by $L+\hat{x}$, where $\hat{x}$ represents the position operator of sample.

\begin{figure}[h]
\includegraphics[scale=0.45]{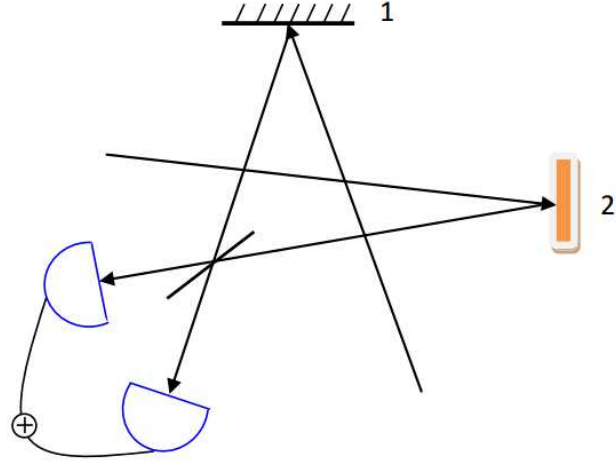}
 \caption{\label{fig.1}Schematic diagram of modified Michelson interferometer. The mirror 2 is replaced by the sample with temperature $T$. The information of temperature is carried by the reflected photons, which are measured at the output ports. }
 \end{figure}
We consider a simple harmonic oscillator model.  The Hamiltonian of  sample is simply described by
\begin{equation}
H=\hbar\omega a^\dagger a,
\end{equation}
where $\omega$ represents the vibrational frequency, $a^\dagger(a)$ denotes the creation (annihilation) operator.
The position operator $\hat{x}$ can be described by
\begin{equation}
\hat{x}=\sqrt{\frac{\hbar}{2m\omega}}(a^\dagger+a),
\end{equation}
where $m$ denotes the mass of harmonic oscillator (sample).
The sample is in the thermalized mixed state and the corresponding density matrix is given by
\begin{equation}
\rho=\exp(-\beta H)/Tr[\exp(-\beta H)],
\end{equation}
with $\beta=1/(KT)$ where $K$ denotes the Boltzmann constant and $T$ denotes the temperature of sample that we want to obtain.

The Hamiltonian of photons in the interferometer is given by
\begin{equation}
H_j=\hbar\omega_pA_j^\dagger A_j,
\end{equation}
where $\omega_p$ represents the frequency of input photons, $j=1,2$ denotes two arms.

The output photons are connected with the input photons by the unitary operator
  \begin{equation}
  U=U_1U_2
\end{equation}
in which,
 \begin{equation}
  U_j=\exp[-iH_jL_j/(\hbar c)],
\end{equation}
where $c$ represents the propagation speed of light in vacuum.
 When the input state is a coherent state$|\alpha/\sqrt{2}\rangle_1|\alpha/\sqrt{2}\rangle_2$, the output state is $U_1U_2|\alpha/\sqrt{2}\rangle_1|\alpha/\sqrt{2}\rangle_2$ with $\alpha=\sqrt{N}$.

 The measurement is performed by registering the sum in the number of photons recorded by two detectors at the output ports of the modified interferometer. The corresponding measurement operator can be expressed as $M=(A_2^\dagger A_1+A_1^\dagger A_2)$.

\section{measurement of temperature and the resolution}
Due to the thermal fluctuation of sample, the expectation value of observed quantity $M$ changes as follows:

 \begin{equation}
 \langle M\rangle=Tr[\rho N\cos(\omega_p\hat{x}/c)]
\end{equation}
Substituting the thermalized mixed state of sample Eq.(3) into the above equation, we get
 \begin{equation}
 \langle M\rangle=Tr[\exp(-\beta H) N\cos(\omega_p\hat{x}/c)]/Tr[\exp(-\beta H)].
\end{equation}
In general the thermal fluctuation is weak, we suppose that  $\langle(\omega_p\hat{x}/c)^2\rangle\ll1$.
For better analysis of the question, we consider two situations: $\frac{\hbar\omega}{KT}\ll1$ and $\frac{\hbar\omega}{KT}\gg1$.

In the case $\frac{\hbar\omega}{KT}\ll1$, the energy level spectrum of sample can be treated as a continuous distribution.
We can obtain the expectation value of the observed M (see the detail calculation in Appendix A)
\begin{eqnarray}
\langle M\rangle
=N-N(\omega_p/c)^2\frac{\hbar}{4m\omega}(1+2\frac{1}{\beta\hbar \omega}).
\end{eqnarray}
So the temperature of sample $T$ is related to the observed $\langle M\rangle$ as follows:
\begin{eqnarray}
T=\frac{\langle M\rangle-N+N(\omega_p/c)^2\frac{\hbar}{4m\omega}}{N(\omega_p/c)^2\frac{K}{2m\omega^2}}.
\end{eqnarray}

In the case $\frac{\hbar\omega}{KT}\gg1$, the contribution from high energy level can be neglected.
We just consider the two lowest levels.
 \begin{eqnarray}
\langle M\rangle
\simeq N-\frac{N\Sigma_{n=0}^1\exp[-n\beta\hbar \omega](\omega_p/c)^2(2n+1)\frac{\hbar}{4m\omega}}{\Sigma_{n=0}^1\exp[-n\beta\hbar \omega]}.
\end{eqnarray}
The temperature of sample can be described by
\begin{eqnarray}
T=\frac{\hbar\omega}{K\ln\frac{3b-d}{d-b}}
\end{eqnarray}
with $b=(\omega_p/c)^2\frac{\hbar}{4m\omega}$ and $d=\frac{N-\langle M\rangle}{N}$.

Next, we explore the measurement resolution of temperature. Due to that continuous spectrum is common\cite{lab8,lab9,lab10}, we just analyze the measurement resolution of temperature in the case  $\frac{\hbar\omega}{KT}\ll1$. The resolution of temperature can be estimated by noise-to-signal ratio
\begin{eqnarray}
\delta T=\frac{(\Delta M)^2}{|\frac{d\langle M\rangle}{dT}|^2}
\end{eqnarray}
The uncertainty $\triangle M\simeq N$, which is calculated in Appendix B.
As a result, the resolution of temperature is obtained
 \begin{eqnarray}
\delta T=\frac{2m\omega^2}{NK(\omega_p/c)^2}.
\end{eqnarray}
From this result, we can see that it satisfies the quantum limit. The resolution is proportional to $1/N$. And increasing the photon-to-sample frequency ratio $\omega_p/\omega$, the resolution can be enhanced.
\section{Two thermalized mirrors}
In this section, we explore that both of the two reflected mirrors are replaced by two same thermalized samples. The mirror 1 in Fig.1 is also replaced by a thermalized mirror (sample).
So the length of two arms can be expressed by $L_j=L+\hat{x}_j$ with $j=1,2$. The position operator of sample $j$ can be described by the corresponding creation and annihilation operator, $\hat{x}_j=\sqrt{\frac{\hbar}{2m\omega}}(a_j^\dagger+a_j)$.

Using the way in section II, we can get the expectation value of observed M
 \begin{equation}
 \langle M\rangle=Tr[\rho_1\rho_2 N\cos(\omega_p(\hat{x}_1-\hat{x}_2)/c)]
\end{equation}
where $\rho_j$ represents the density matrix of sample $j$.

When we just calculate up to the second-order term, the above equation can be simplified as follows:
 \begin{equation}
 \langle M\rangle=N-2Tr[\rho_1 N\cos(\omega_p\hat{x}_1/c)].
\end{equation}
Considering the continuous spectrum, the expectation value  of M can be achieved
 \begin{equation}
 \langle M\rangle=N-2N(\omega_p/c)^2\frac{\hbar}{4m\omega}(1+2\frac{1}{\beta\hbar \omega}).
\end{equation}
Following that, the temperature is obtained
\begin{eqnarray}
T=\frac{\langle M\rangle-N+N(\omega_p/c)^2\frac{\hbar}{2m\omega}}{N(\omega_p/c)^2\frac{K}{m\omega^2}}.
\end{eqnarray}
Using the method in Appendix B, the resolution of temperature is
 \begin{eqnarray}
\delta T=\frac{m\omega^2}{NK(\omega_p/c)^2}.
\end{eqnarray}
Obviously, the resolution is improved up to twice than the former resolution in Eq.(14). It show that using two thermalized samples is useful in enhancing the resolution of temperature. Because ultimately, the more information resources available to us the better.
\section{Nonlinear medium}
In order to improve the measurement precision of temperature, we hope to employ new resources. In this setup the entanglement or squeezed state can not improve the precision of temperature, which can be verify easily. So nonlinear resource is a possible choice.
We assume that the modified interferometer is embedded in a gas with Kerr nonlinearity.

In this nonlinear gas, the Hamiltonian of photons in the interferometer should be modified as follows:
\begin{equation}
H_j=\hbar\omega_p[A_j^\dagger A_j+\frac{\chi}{2}(A_j^\dagger A_j)^2],
\end{equation}
where $\chi$ denotes the nonlinear phase shift per photon\cite{lab6}. The corresponding unitary operator is described by
\begin{equation}
  U_j=\exp[-in_0H_jL_j/(\hbar c)],
\end{equation}
where $n_0$ represents the refractive index of gas that decides the speed of light.

In this section, we only consider one thermalized sample. When involving the nonlinear contribution, we find the observed $M$ can be obtained by replacing $\omega_p$ with $\omega_p'$ (the detail result in Appendix C).
\begin{eqnarray}
\langle M\rangle=N-N(\omega_p'/c)^2\frac{\hbar}{4m\omega}(1+2\frac{1}{\beta\hbar \omega})
\end{eqnarray}
where $\omega_p'=[1+\frac{\chi}{2}(N+1)]n_0\omega_p$. Due to the nonlinear gas, the expectation value of observed M is obviously changed.
Then the temperature is expressed by
\begin{eqnarray}
T=\frac{\langle M\rangle-N+N(\omega_p'/c)^2\frac{\hbar}{2m\omega}}{N(\omega_p'/c)^2\frac{K}{m\omega^2}}.
\end{eqnarray}
Due to the nonlinear constant $\chi\ll1$\cite{lab11,lab12,lab13}, the expectation value of $M^2$ is still approximate to $N$.
So the resolution of temperature is achieved
 \begin{eqnarray}
\delta T=\frac{m\omega^2}{NK(\omega_p'/c)^2}.
\end{eqnarray}
When $N\chi\gg1$, $\omega_p'\gg\omega_p$. Namely, the resolution of temperature can be significantly improved due to the nonlinear gas.
\section{Discussion}
The main adverse factors are the efficiency of detection $\eta_1$ and the reflection efficiency of sample $\eta_2$.
The resolution of temperature will be decreased. We can obtain the corresponding uncertainty of temperature by supposing that $A_j$ is replaced by $A_j=\sqrt{\eta_1\eta_2}A_j+\sqrt{1-\eta_1\eta_2}B_j$. $B_j$ is uncorrelated field mode. As a result, the uncertainty of temperature is given by
 \begin{eqnarray}
\delta T=\frac{m\omega^2}{\eta_1\eta_2NK(\omega_p'/c)^2}.
\end{eqnarray}

Next, we choose a set of parameters which can be performed in experiment.
The nonlinear constant $\chi$ can arrive at $10^{-8}$\cite{lab12}.  The total number of input visible photons is given: $N=10^{10}$.
The frequency of photon $\omega_p=10^{10}Hz$. The temperature of sample is chosen as: $KT\simeq10^{-21}J$. For the sample, the corresponding parameters are given: $m=10^{-10}Kg$, $\omega=10^2Hz$. Consider the perfect efficiency of detection and reflection:
$\eta_1=\eta_1=1$. In general, the refractive index of gas is close to 1: $n_0\approx1$. Substituting those parameters into Eq.(25), the final resolution of temperature is figured out: $\delta T\approx10^{-7}\mathbf{K}$, where $\mathbf{K}$ denotes the Kelvin units of temperature. The corresponding relative uncertainty is obtained: $\frac{\delta T}{T}\simeq10^{-9}$. The relative uncertainty of the Boltzmann constant is $1.7*10^{-6}$ from CODATA\cite{lab14}.
So using the modified interferometer embedded in a nonlinear gas can obtain enough high resolution of temperature with no need for the direct measurement of population. It lays the foundation of obtaining more precise Boltzmann constant.

In this article, we only consider the sample consisted of single harmonic oscillator. It can be generalized to a more complex sample.
The encoding way of temperature will be changed. The thermal fluctuation can induces the uncertainty of position that photons are reflected. The wave function of sample can be expressed as $\psi(x)=\frac{1}{\sqrt{2\pi}\Delta^2(T)}\exp[-\frac{(x-x_0)^2}{\Delta^2(T)}]$. The information of temperature is encoded into $ \Delta(T)$. In another word, for a more complex case,  finding the form of $ \Delta(T)$ is a key step. This depends on the detail sample.
\section{Conclusion}
In this article, we investigate that a modified Michelson can measure the temperature. The information of temperature is encoded into the reflected photons. So it does not need to measure the population of sample directly. Both of two mirrors replaced by two same thermalized samples performs twice than only one mirror replaced by the corresponding thermalized sample. For significantly improving the resolution of temperature, the nonlinear resource is required. The modified interferometer is embedded in a nonlinear gas, which displays Kerr nonlinearity. As a result, we prove that the nonlinearity can enhance the resolution of temperature obviously.

We believe that the scheme will help to set up a new accurate thermometer in experiment. A more complex sample deserves the further development. The exploration of sample's structure is a key step.

\section*{Acknowledgement}
This work was supported by the National Natural Science Foundation of China under Grant  No. 11375168.

\newpage
\textbf{Appendix A: Expectation of the observed M}\\
The statistics of the observed M is expressed as
\begin{equation}
\langle M\rangle=Tr[\rho\langle\alpha/\sqrt{2}|_1\langle\alpha/\sqrt{2}|_2U_1^\dagger U_2^\dagger MU_1U_2|\alpha/\sqrt{2}\rangle_1|\alpha/\sqrt{2}\rangle_2].
\end{equation}
Utilizing the relation $AF(A^\dagger A)=F(A^\dagger A+1)A$, we can obtain
\begin{equation}
U_j^\dagger A_jU_j=\exp[i\omega_pA_j^\dagger A_jL_j/c]A_j\exp[-i\omega_pA_j^\dagger A_jL_j/c]\
=\exp[-i\omega_pL_j/c]A_j.
\end{equation}
Combining the above equations,  we arrive at
\begin{equation}
\langle M\rangle=Tr[\rho N\cos(\omega_p\hat{x}/c)]
\end{equation}
Note that $\langle(\omega_p\hat{x}/c)^2\rangle\ll1$,  $\cos(\omega_p\hat{x}/c)\approx1-1/2(\omega_p\hat{x}/c)^2$

Using the eigenstate of the Hamiltonian of sample,
\begin{equation}
\langle n|\hat{x}^2|n\rangle=\langle n|\frac{\hbar}{2m\omega}(a^\dagger+a)^2|n\rangle\\
=(2n+1)\frac{\hbar}{2m\omega}
\end{equation}

 For the case $\frac{\hbar\omega}{KT}\ll1$, $Tr[\exp[-\beta H]]\simeq\frac{1}{\beta\hbar \omega}$.
 \begin{eqnarray}
\langle M\rangle=Tr[\rho N\cos(\omega_p\hat{x}/c)]\beta\hbar \omega
=N-Tr[\exp[-\beta H] N(\omega_p\hat{x}/c)^2]\beta\hbar \omega\\
=N-N/2\Sigma_{n=0}^\infty\exp[-n\beta\hbar \omega](\omega_p/c)^2(2n+1)\frac{\hbar}{2m\omega}\beta\hbar \omega\\
\simeq N-N/2\int_{n=0}^\infty\exp[-n\beta\hbar \omega](\omega_p/c)^2(2n+1)\frac{\hbar}{2m\omega}\beta\hbar \omega\\
=N-N(\omega_p/c)^2\frac{\hbar}{4m\omega}(1+2\frac{1}{\beta\hbar \omega})
\end{eqnarray}

 For the case $\frac{\hbar\omega}{KT}\gg1$,
 \begin{eqnarray}
\langle M\rangle=Tr[\rho N\omega_p\cos(\hat{x}/c)]\beta\hbar \omega\\
=N-N/2\Sigma_{n=0}^\infty\exp[-n\beta\hbar \omega](\omega_p/c)^2(2n+1)\frac{\hbar}{2m\omega}/\Sigma_{n=0}^\infty\exp[-n\beta\hbar \omega]\\
\simeq N-\frac{N\Sigma_{n=0}^1\exp[-n\beta\hbar \omega](\omega_p/c)^2(2n+1)\frac{\hbar}{4m\omega}}{\Sigma_{n=0}^1\exp[-n\beta\hbar \omega]}\\
\end{eqnarray}
\textbf{Appendix B: uncertainty of the observed M}\\

\begin{eqnarray}
 \langle M^2\rangle=\langle {A_2^\dagger}^2 A_1^2+{A_1^\dagger}^2 A_2^2+A_2^\dagger A_2A_1A_1^\dagger+A_1^\dagger A_1A_2A_2^\dagger \rangle.
\end{eqnarray}
Utilizing the relation $A^2F(A^\dagger A)=F(A^\dagger A+2)A^2$, the above equation can be calculated as
\begin{eqnarray}
\langle M^2\rangle=\langle N^2/2\cos(2\omega_p\hat{x}/c)+N^2/2+N\rangle\\
\simeq N^2+N-N^2(\omega_p/c)^2\langle\hat{x}^2\rangle\\
=N^2+N-N^2(\omega_p/c)^2\frac{\hbar}{2m\omega}(1+2\frac{1}{\beta\hbar \omega})N^2(\omega_p/c)^2
\end{eqnarray}
Combining Eq.(41)and Eq.(33),
we get the uncertainty of the observed M
\begin{eqnarray}
\Delta M\approx N.
\end{eqnarray}
\begin{eqnarray}
\frac{d\langle M\rangle}{dT}=-N(\omega_p/c)^2\frac{K}{2m\omega^2}
\end{eqnarray}
So the uncertainty of temperature is
 \begin{eqnarray}
\delta T=\frac{2m\omega^2}{NK(\omega_p/c)^2}
\end{eqnarray}

\textbf{Appendix C: Nonlinear medium}\\
When the interferometer is embedded in a nonlinear gas, the corresponding unitary transformation is different.
\begin{eqnarray}
U_j^\dagger A_jU_j=\exp[i(\omega_pA_j^\dagger A_j+\frac{\chi}{2}(A_j^\dagger A_j)^2)n_0L_j/c]A_j\exp[-i(\omega_pA_j^\dagger A_j+\frac{\chi}{2}(A_j^\dagger A_j)^2)n_0L_j/c]\\
=\exp[-i\omega_pn_0L_j/c(1+\frac{\chi}{2}(2A_j^\dagger A_j+1))]A_j.
\end{eqnarray}
The expectation of observed M can be calculated
\begin{eqnarray}
 \langle M\rangle=Tr[\rho\langle\alpha/\sqrt{2}|_1\langle\alpha/\sqrt{2}|_2U_1^\dagger U_2^\dagger MU_1U_2|\alpha/\sqrt{2}\rangle_1|\alpha/\sqrt{2}\rangle_2]\\
 =Tr[\rho N\cos[(1+\frac{\chi}{2}(N+1))n_0\omega_p\hat{x}/c]]
\end{eqnarray}
Comparing with Eq.(28), it can be treated as $\omega_p$ replaced by $[1+\frac{\chi}{2}(N+1)]n_0\omega_p$.
Considering continuous spectrum, the expectation is obtained
\begin{eqnarray}
\langle M\rangle=N-N[(1+\frac{\chi}{2}(N+1))n_0\omega_p/c]^2\frac{\hbar}{4m\omega}(1+2\frac{1}{\beta\hbar \omega})
\end{eqnarray}
\begin{eqnarray}
\langle M^2\rangle=\langle N^2/2\cos(2\omega_p\hat{x}/c)+N^2/2+N\rangle\\
=N^2+N-N^2[(1+\frac{\chi}{2}(N+2))n_0\omega_p/c]^2\frac{\hbar}{2m\omega}(1+2\frac{1}{\beta\hbar \omega})N^2(\omega_p/c)^2
\end{eqnarray}

Due to the nonlinear constant $\chi\ll1$, we can obtain the same uncertainty of M
\begin{eqnarray}
\Delta M\approx N.
\end{eqnarray}

\end{document}